\documentclass[aps,prl,reprint,groupedaddress]{revtex4-1}
\usepackage{amsmath}
\usepackage{amssymb}
\usepackage{mathrsfs}
\usepackage{amsthm}
\usepackage{chemarrow}
\usepackage{times}
\usepackage{xcolor}

\begin{document}
\def\rd{{\rm d}}
\def\vf {{\bf f}}
\def\vg {{\bf g}}
\def\vh {{\bf h}}
\def\vj {{\bf j}}
\def\vP{{\bf p}}
\def\vp{{\bf p}}
\def\vx{{\bf x}}
\def\vy{{\bf y}}
\def\vY{{\bf Y}}
\def\vz{{\bf z}}
\def\vn{{\bf n}}
\def\vu{{\bf u}}
\def\vv{{\bf v}}
\def\vw {{\bf w}}
\def\mA{{\bf A}}
\def\mB{{\bf B}}
\def\mC{{\bf C}}
\def\mG{{\bf G}}
\def\mI{{\bf I}}
\def\mS{{\bf S}}
\def\mQ{{\bf Q}}
\def\mR{{\bf R}}
\def\mT{{\bf T}}
\def\mU{{\bf U}}
\def\mV{{\bf V}} 

\def\mGa{\mbox{\boldmath$\Gamma$}}
\def\mPhi{\mbox{\boldmath$\Phi$}}
\def\mPi{\mbox{\boldmath$\Pi$}}
\def\mXi{\mbox{\boldmath$\Xi$}}
\def\vbeta {\mbox{\boldmath$\beta$}}
\def\veta {\mbox{\boldmath$\eta$}}
\def\vepsilon{\mbox{\boldmath$\varepsilon$}}
\def\vgamma {\mbox{\boldmath$\gamma$}}
\def\vxi {\mbox{\boldmath$\xi$}}
\def\vpi{\mbox{\boldmath$\pi$}}

\newtheorem{theorem}{Theorem}[section]
\newtheorem{lemma}[theorem]{Lemma}
\newtheorem{corollary}{Corollary}[theorem]

\title{Emergence and Breaking of Duality Symmetry in Generalized Fundamental Thermodynamic Relations
%:\\ Repeated Measurements and 
%Macroscopic Limit
}

\author{Zhiyue Lu}
\email{zhiyuelu@unc.edu}
\address{Department of Chemistry,
University of North Carolina,
Chapel Hill, NC 27599-3290, U.S.A.}
\author{Hong Qian}
\email{hqian@u.washington.edu}
\address{Department of Applied Mathematics,
University of Washington,
Seattle, WA 98195-3925, U.S.A.
}

\begin{abstract}
Thermodynamics as limiting behaviors of statistics is generalized to arbitrary system with probability {\it a priori} where thermodynamic infinite-size limit is replaced by multiple-measurement limit. A duality symmetry between Massieu's and Gibbs' entropy arises in the limit of infinitely repeated observations, yielding the Gibbs equation and Hill-Gibbs-Duhem equation (HGDE) as dual pair. If a system has thermodynamic limit satisfying Callen's postulate, entropy being an Eulerian function, the symmetry is lost: the HGDE reduces to the Gibbs-Duhem equation. This theory provides a de-mechanized foundation for classical and nanothermodynamics and offers a framework for distilling emergence from large data, free from underlying details.\\

\end{abstract}

\baselineskip = 0.19 in

\maketitle
For macroscopic equilibrium systems approaching infinite-size limit, classical thermodynamics emerges systematically in Gibbs's ensemble theory of statistical mechanics \cite{callen-book,guggenheim1933modern,fermi1956thermodynamics}. It states that equilibrated macroscopic homogeneous systems (i.e., simple systems) can be fully described by a number of extensive quantities $S,U,V,N$ that are related by the fundamental relation $U=U(S,V,N)$ or equivalently $S=S(U,V,N)$. All other thermodynamic properties, such as temperature, pressure, heat capacity, and Helmholtz free energy, can be obtained through partial derivatives or Legendre transforms of the fundamental equation. Moreover, there exist a set of universal relations despite the system's specific details entailed by the particular functional form: Based upon the extensivity of $S,U,V,N$ and Euler's theorem for homogeneous degree one functions (\ref{eqn:Euler_eqn}), one obtains the Gibbs-Duhem equation as a differential form (\ref{eqn:GD_eqn}) of the universal relation \cite{callen-book}. Taking the entropy representation of fundamental relation in a compact vector representation, the two relations are
\begin{eqnarray}
	  && S=\vbeta \cdot \vY,
\label{eqn:Euler_eqn}
\\
	  && 0 =  \vY  \cdot {\rm d} \vbeta ,
\label{eqn:GD_eqn}
\end{eqnarray}
where $\vbeta = (1/T, P/T, -\mu/T)$ and $\vY=(U,V,N)$.

For small systems without the thermodynamic limit, in contrast, $S(U,V,N)$ is not an Eulerian function of $U,V,N$ and the Gibbs-Duhem equation is no longer valid. Moreover, proper definitions of $U,V,S,N$ become non-trivial: They become dependent upon the methods for measurements due to fluctuations. To recover thermodynamic-like principles for small systems, there have been two main approaches: One carefully redefines thermodynamic quantities for small systems (e.g., volume) that satisfy relations that preserve the from of macroscopic thermodynamics.  The recent work of Seifert \cite{seifert2016first} and Jarzynski \cite{jarzynski2017stochastic} has taken up
this bottom-up approach. Another approach developed by T. L. Hill, more ``phenomenological and thermodynamic'' in nature, introduced additive observables of a small system as extensive quantities and presented thermodynamic relations with a finite-size correction term to the entropy function \cite{hill-nano,kjelstrup2021nanoscale,bedeaux-cpl,bedeaux}. The latter theory evoked a replica averaging technique and obtained {\it modified Gibbs-Duhem equation} through Legendre transform of non-Eulerian entropy function:  
\begin{eqnarray} 
	  && S= \vbeta \cdot \vY-\frac{\mathcal{E}}{T},
\label{eqn:EulerHill_eqn}
\\
	  && {\rm d} \left (\frac{\mathcal{E}}{T} \right) =  \vY  \cdot {\rm d} \vbeta.
\label{eqn:GDH_eqn}
\end{eqnarray}
The relations (\ref{eqn:EulerHill_eqn}) and (\ref{eqn:GDH_eqn}) are general; one natural example for $\mathcal{E}$ is the surface-effect correction to bulk thermodynamics; see \cite{bedeaux-cpl} for a recent treatment of this problem.  In Hill's nanothermodynamics, which was inspired by ideal solution where solutes are replica of small systems, the thermodynamic limit is replaced by a {\it replica ensemble limit}. 

Although the mathematical foundation of classical thermodynamics emerging from the macroscopic limit has been extensively studied in statistical mechanics, neither the theoretical nor probabilistic foundation of Hill's nanothermodynamics is well understood. In this letter, we explicitly discuss the concept of the mean values of repeated measurements as mesoscopic thermodynamic quantities and a general framework that includes both Gibbs's ensemble and Hill's replica approaches. We also demonstrate that Hill's replica approach can be extended to arbitrary stochastic systems with certain stationarity, for examples translational invariance in either space or time. This setting allows for applying the large deviation principle from the theory of probability, upon which we derive a set of relationships that captures and generalizes Hill's thermodynamic relations. We find that the generalized fundamental equation obtained through the large deviation approach exhibits a duality symmetry under a full, Legendre-Fenchel transform (LFT). According to the large deviation theory, the LFT based on variational calculus (as in Eq. \ref{lft}) rather than the traditional Legendre transform based on derivative is critical for recognizing this duality, as echoed in a recent study \cite{galteland}.
  The duality symmetry is apparently broken in the classical macroscopic thermodynamics theory. 

With the generalized thermodynamic framework, when one takes the thermodynamic limit, the symmetry broken can be explicitly illustrated as a sub-extensive term. Our result shows the large deviation theory as the mathematical foundation to the emergence of universal thermodynamic-like behavior for arbitrary stationary systems with or without detailed balance condition, in terms of generalized physical quantities not limited to the traditional $S,U,V,N$. 

In classical thermodynamics \cite{callen-book} the fundamental relation $S=S(U,V,N)$, e.g., entropy and other thermodynamic potentials as Eulerian degree-one functions of the extensive quantities, is an asymptotic limiting behavior of statistics \cite{jaynes1957information,chibbaro-et-al,qct}.  The mathematical theory of large deviations in probability provides not only powerful tools for statistical mechanics \cite{ellis-book,oono,touchette2009large,ge-qian-ijmpb,dembo-book}, but more importantly a novel organizational principle for the theory of thermodynamic behavior itself \cite{smith}, in particular the origin of {\em maximum entropy principle} \cite{dill,qian-cheng-qb} as a consequence of contraction principle and Gibbs conditioning \cite{dembo-book}. To take advantage of large deviation principle in an arbitrary stationary system of an arbitrary size, we explicitly discuss the {\it repeated measurement limit} inspired by Hill's work \cite{hill-nano,hill-book}. Consider a general stochastic system with state space $\mathfrak{S}$ and {\em a priori} probability density function $f_{\vx}(\vx)$, $\vx\in\mathfrak{S}$, the large deviations theory (LDT) states that for a large number $M$ repeated measurements of an 
array of $K$ real observables 
$\vg(\vx)=\big(g_1, \cdots,g_K\big)(\vx)$, one can construct
the mean value
\begin{equation}
   \overline{\vg}_M = \frac{\vg^{(1)}+\cdots+\vg^{(M)}}{M},
\end{equation}
in which the vector $\vg^{(i)}$ is the $i^{th}$ measurement outcome and $\overline{\vg}_M$ is the mean or the $M$ repeated measurements. For a stationary stochastic system, we can treat each $\vg^{(i)}$ as identically 
distributed  random variables. As $M\to\infty$, the law of large number expects $\overline{\vg}_M$ converges to the expectation value of $\vg(\vx)$ defined as:
\begin{equation}
 \mathbb{E}[\vg] = \int_{\mathfrak{S}} f_{\vx}(\vx) \vg(\vx) \rd\vx.
\end{equation}
However, when $M\gg 1$ is large and finite, the LDT predicts that the probability distribution
for the measurement mean $\overline{\vg}_M$ follows an asymptotic expression, 
\begin{equation}
    	\ln \Pr\big\{ \vy < \overline{\vg}_M \le \vy+\rd\vy\big\} 
      =  M\eta(\vy) + o(M),
\label{massieuent}
\end{equation}
where the negative large deviation rate function $\eta(\vy)$ is a generalized entropy function, the term $o(M)$ is vanishing small compared to $M\eta(\vy)$ in the repeated measurement limit $M\to \infty$. According to Cram\'{e}r's theorem \cite{ellis-book,dembo-book,touchette2009large} for independent $\vg^{(i)}$'s, one can also obtain a pair of conjugate functions
\begin{eqnarray}
	  && \eta(\vy) = \min_{\vbeta} \left\{ \vbeta\cdot\vy + \psi(\vbeta)
       \right\},
\label{lft}
\\
	 && \psi(\vbeta) = \ln \int_{\mathfrak{S}} f_{\vx}(\vx) 
        \exp\big(-\vbeta\cdot
          \vg(\vx)\big)\rd\vx,
\label{pf}
\end{eqnarray}
in which $\vy=(y_1,\cdots,y_K)$ are dummy variables representing the possible values that the $K$ measurement means could take, and $\vbeta=(\beta_1,\cdots,\beta_K)$ serves as their {\it conjugates}.  See \cite{qian-cheng-qb} for a highly condensed but self-contained tutorial of the large deviation analysis when $K=1$. In this letter we will show that $\eta(\vy)$ is analogous to Hill's modified thermodynamic fundamental equation $S=S(U,V,N)$ and $\psi(\vbeta)$ is an sub-extensive term that is present for small systems but negligible in the thermodynamic limit. Notice that the {\it repeated measurement limit}, $M\rightarrow \infty$, needs to be distinguished from the thermodynamic limit where a system's size approaches infinity. The repeated measurement limit allows us to apply large deviation theory to systems of arbitrary size. 

{\bf\em Reduction to classical thermodynamics and statistical mechanics.}
Although the proposed multiple-measurement approach can be applied to arbitrary stationary system, equlibrium or non-equilibrium, or even a-thermal (e.g., systems very far removed from a molecular understanding and the possibility of a Hamiltonian conception, such as those in ecology), the large deviation theory analysis bears a unmistakable resemblance to those in statistical mechanics. We emphasize that the various ensembles originating from the microcanonical ensemble in statistical thermodynamics are achieved by partially replacing a subset of the components of $\vg$ by it's conjugate in $\vbeta$ in (\ref{lft}) and (\ref{pf}), which we denote as partial LFTs and will be discussed elsewhere.  Here we focus on the {\it full LFT}, replacing all of $\vy$ by $\vbeta$. For illustrative purposes, let us identify $\vx$ as the phase space point, and equal {\em a priori} measure for $\vx\in\mathfrak{S}$, and perform a scalar measurement ($K=1$)  $g(\vx)=\mathcal H(\vx)$ as the mechanical energy (Hamiltonian). Then $\overline g(\vx)$ is statistical sampling mean of internal energy and the integral in (\ref{pf}) is canonical partition function. Moreover, $-\beta^{-1}\psi(\beta)$ is the Helmholtz free energy and $k_B\psi(\beta)$ is known as Massieu-Planck free entropy \cite{planck-book}.
More interestingly, noting that $\psi(\beta)$ is a convex function, 
the LFT in (\ref{lft}) can be carried 
out with differentiation:
\begin{equation}
     \eta(y) = \left[-\beta\left(\frac{\partial\psi}{\partial\beta}\right) 
                + \psi(\beta) \right]_{\beta:\partial\psi/\partial\beta=-y}
	= \frac{\partial [\beta^{-1}\psi(\beta)]}{\partial (1/\beta)}.
\label{gibbsent}
\end{equation}
This shows that $k_B\eta(y)$ corresponds to the thermodynamic entropy as a function of internal energy $S(U)$.

To make a more precise analogy to the thermodynamic fundamental relation $S=S(U,V,N)$, let us consider the $K=3$ scenario where measurable quantities in $\vg(\vx)$ represents the Hamiltonian, volume, and number of particles of an ergodic mechanical system that satisfies detailed balance condition ($\vy=(U,V,N)$), then the integral in (\ref{pf}), as a function of $\vy$'s conjugate variables $\vbeta=(k_BT)^{-1}(1,P,-\mu)$, corresponds to a special partition function for the $\vbeta$-ensemble. This $\vbeta$-ensemble partition function is not frequently discussed in the literature (see Guggenheim's work \cite{gugg-1939}), but could be related to the full Legendre transform of $S(U,V,N)$: If one perform the full Legendre transform of the fundamental relation $S=S(U,V,N)$ in the classical thermodynamic theory, one apparently expects to obtain $0$. However, by performing a statistical mechanical analysis of the $\vbeta$-ensemble, the corresponding partition function $\exp\big(\psi(\vbeta)\big)$ becomes the so-called Guggenheim's generalized partition function\cite{gugg-1939}. For macroscopic thermodynamics, this corresponding thermodynamic potential is ignored because it is sub-extensive.\footnote{We discover that the spirit of Guggenheim's setup was similar to J. G. Kirkwood's introduction of the {\em potential of mean force}, in {\em J. Chem. Phys.} {\bf 3}, 305--313, (1935).} In nanothermodynamics, however, the free energy like term $-k_BT\psi$ is named {\em subdivision potential} by Hill \cite{hill-book}. Although our approach can be applied to non-detail-balanced systems and the choices of measurable quantities are not limited by thermodynamics ($U,V,N$), we borrow the names from thermodynamics and name $\psi$ the Massieu-Guggenheim free entropy and correspondingly $\eta$ the Gibbs entropy. For simplicity, we will denote them as free entropy and entropy in the rest of the letter.

{\bf\em Free entropy--Gibbs entropy duality symmetry.} It is always important to identify a thermodynamic potential function together with its appropriate independent variables in thermodynamics.  The mathematics in equations (\ref{massieuent}-\ref{gibbsent}) 
reproduces some of the key steps associated with the changing from one set of variables (ensemble) to another in statistical mechanics. In thermodynamics, such change of ensemble are represented by Legendre transformations.  More importantly, the large deviation rate function $\eta(\vy)$ in our general approach is a convex function of $\vy$ since it is the LFT of the convex function $\psi(\vbeta)$ as a function of $\vbeta$, and (\ref{lft}) has an inverse 
relation \cite{rockafellar}:
\begin{equation}
   \psi(\vbeta) = \max_{\vy}\Big\{-\vbeta\cdot\vy+\eta(\vy)\Big\}.
\label{massieu}
\end{equation}
Even though we first obtain $\psi(\vbeta)$ from $f_{\vx}(\vx)$ 
and $\vg(\vx)$ through the partition function, some detailed information concerning the original system is lost in the statistics
of the mean value \footnote{One of course could construct 
mean square value and many other nonlinear statistics to gain
additional information.  The relationship between such an approach and traditional thermodynamics remains to be elucidated.}.  Transcending the original stochastic system, the free entropy $\psi(\vbeta)$ and entropy $\eta(\vy)$ 
in (\ref{lft}) and (\ref{massieu}) now form a dual under LFT: A duality symmetry emerges!
This symmetry can be expressed as the generalized Euler's equation with Hill's correction for the entropy or the free entropy:
\begin{subequations}
\begin{eqnarray}
		\eta(\vy) &=& \vbeta\cdot\vy + \psi(\vbeta),
\label{eqn: Euler-Hill-general1}\\
	   \psi(\vbeta) &=& -\vy\cdot\vbeta + \eta(\vy).
\label{eqn: Euler-Hill-general2}
\end{eqnarray}
\end{subequations}
where $\vbeta$ is the gradient of $\eta(\vy)$ and $\vy$ is the gradient of $\psi(\vbeta)$. Notice that our equation \ref{eqn: Euler-Hill-general1} extends Hill's equation
\begin{equation}
    S=\frac{U}{T} + \frac{PV}{T} - \frac{\mu N}{T} - \frac{\mathcal{E} }{T}
\end{equation}
to be applicable for arbitrary stationary stochastic systems. 
One can also express the symmetry in terms of the exact differentials:
\begin{subequations}
\begin{eqnarray}
		\rd\eta(\vy) &=& \vbeta\cdot\rd\vy,
\label{ftr}\\
	   \rd\psi(\vbeta) &=& -\vy\cdot\rd\vbeta.
\label{hgde}
\end{eqnarray}
\end{subequations}
If we again identify $\vy$ as $(U,V,N)$ in statistical 
mechanics, $k_B\eta(\vy)$ as entropy $S(U,V,N)$, and correspondingly 
$\vbeta=(k_BT)^{-1}(1,P,-\mu)$, we see that (\ref{ftr}) 
is a generalization of the Gibbs equation 
\begin{subequations}
\label{dual-rel}
\begin{equation}
          \rd S = \left(\frac{1}{T}\right)\rd U + \left(\frac{P}{T}\right)\rd V - \left(\frac{\mu}{T}\right) \rd N
\end{equation} 
which holds for both Hill's nanothermodynamics and classical macroscopic thermodynamics. Moreover, our equation (\ref{hgde}) generalizes a key result of Hill's nanothermodynamics \cite{hill-nano,chamberlin2015big}, in which we can find the exact differential of the free entropy:
\begin{equation}
   \rd \left(\frac{\mathcal{E}}{T}\right) = U\rd\left(\frac{1}{T}\right) + V\rd \left(\frac{P}{T}\right) - N\rd\left(\frac{\mu}{T}\right).
\end{equation}
\end{subequations}
Bedeaux et. al. termed relation in (\ref{dual-rel}b) Hill-Gibbs-Duhem equation \cite{bedeaux}.  We shall follow their terminology to call our equation (\ref{hgde}) the generalized Hill-Gibbs-Duhem equation.  One specific example of $\mathcal{E}$ is surface energy, see \cite{bedeaux-cpl} and \cite{dongwei} for recent studies of surface and interface nanothermodynamics. 

These emerging thermodynamic-like results from the multiple-measurement limit presented so far are generally true for any stochastic systems and can be reduced to the description of nanothermodynamics. The symmetry between the generalized entropy $\eta(\vy)$ and generalized Massieu-Planck free entropy $\psi(\vbeta)$, shown by the equations (\ref{eqn: Euler-Hill-general1}), (\ref{eqn: Euler-Hill-general2}), (\ref{ftr}), and (\ref{hgde}), implies a similar symmetry in Hill's nano-thermodynamics: All information about the thermodynamic properties of a given substance, captured by the fundamental equation $S=S(U,V,N)$, is simultaneously contained in Hill's subdivisional function $\mathcal{E}=\mathcal{E}(T,P,\mu)$, which is previously understood as a finite-size correction to the macroscopic thermodynamics. Below, we show that in the thermodynamics limit when the system size approaching infinity, such a symmetry can be broken when one neglects the sub-extensive quantities.

%For a macroscopic system, one additional supposition for $\eta(\vy)$ being an extensive quantity will reduce the l.h.s. of Eq. \ref{hgde} to sub-extensive and effectivly negligible in the thermodynamic limit, resulting in a generalized Gibbs-Duhem equation with $0$ on the l.h.s. 

{\bf\em Broken duality symmetry in the classical thermodynamic limit.}
According to classical thermodynamics, a macroscopic system in the {\em thermodynamic limit} is represented by a set of {\em extensive variables}, which we denote as the vector $\vg$. We adopt Callen's postulate on classical thermodynamics which states that in the thermodynamic limit, entropy being an Eulerian degree-one homogeneous function of all the extensive variables \cite{callen-book}. Thus for a large
system approaching the macroscopic limit, we have  $\vy\to\infty$ and $\eta(\vy)\to\infty$ simultaneously:
\begin{equation}
         \eta(\vy) = y_1\frac{\partial\eta}{\partial y_1}
        + \cdots + y_K\frac{\partial\eta}{\partial y_K}
        + o(\vy)
            = \vy\cdot\vbeta + o(\vy),
\label{macrolim}
\end{equation}
in which the notion $o(\vy)$ represents a sub-extensive term that scales sub-linear with respect to the system's size: $o(\vy)/\vy\to 0$ 
as $\vy\to\infty$. Substituting (\ref{macrolim}) into (\ref{massieu}), one can show that $\psi(\vbeta)$ is purely sub-extensive and disappears in the thermodynamic limit, and thus (\ref{hgde}) is the Gibbs-Duhem equation $\vy\cdot\rd\vbeta=0$ in the (macroscopic) thermodynamic limit. Similarly Hill-Gibbs-Duhem equation (\ref{dual-rel}b) reduces to the Gibbs-Duhem equation
\begin{equation}
     -S\rd T + V\rd P  -N\rd\mu = 0.
\end{equation}
as is expected for classical thermodynamics of macroscopic systems. As a consequence, in the thermodynamic limit for extensive large systems, the duality symmetry
between (\ref{ftr}) and (\ref{hgde}) is apparently lost, if one ignores the sub-extensive term which becomes negligible compared to the extensive terms. When the sub-extensive term is ignored, the full Legendre transformation, which serves as a bridge between the entropy and free entropy becomes illy defined as shown below.

{\bf\em Legendre-Fenchel transformation of a homogeneous function.}
An Eulerian homogeneous function of degree one cannot be a strictly convex function:  If $\eta(\alpha\vy)=\alpha\eta(\vy)$ where $\alpha$ is any real number, then $\eta(\vy)=\vy\cdot\nabla_{\vy}\eta$, and $\nabla_{\vy}\eta = \nabla_{\vy}\eta + \vy\cdot\nabla_{\vy}\nabla_{\vy}\eta$.  This implies
\begin{equation}
      \vy\cdot\nabla_{\vy}\nabla_{\vy}\eta = 0.
      \label{equation12}
\end{equation}
So $\eta(\vy)$ losses strict convexity along the constant $\vy$ direction in which the Hessian matrix $\nabla_{\vy}\nabla_{\vy}\eta$ is singular. The LFT of such an $\eta(\vy)$ can exist only if the $\vy$ is restricted to a compact and convex domain, and the LFT is non-differentiable on the sub-manifold defined by  (\ref{equation12}).\footnote{An illustrative example is $\eta(y_1,y_2)=y_1^2/y_2$ with $y_2\in [b_1,b_2]$.  The corresponding LFT $\psi(\beta_1,\beta_2)= b_1z$ if $z(\beta_1,\beta_2)\triangleq(\beta_2-\beta_1^2/4)>0$, and $=b_2z$ if $z<0$.  The sub-manifold is $z(\beta_1,\beta_2)=0$. When $[b_1,b_2]$ extends to the entire $\mathbb{R}$, $\psi(\beta_1,\beta_2)$ diverges almost everywhere; whose support contracts 
to the sub-manifold.}  When the $\vy$ is extended to the entire $\mathbb{R}^K$, the domain of its LFT is contracted to a $K-1$ sub-manifold which defines the equation of state in classical thermodynamics. Since thermodynamic limit dictates $\eta(\vy)$ being a degree one homogeneous function of $\vy$ \cite{callen-book}, it is the mathematical origin of broken duality symmetry in thermodynamic limit.

{\bf\em Generalized conjugate forces $\vbeta$.}
In thermodynamics $\beta$'s take the forms of $1/T$, $p/T$, and $\mu/T$ and they can be considered as conjugate forces that governs the spontaneous change of energy $U$, volume $V$, and number of particles $N$. We can briefly show that the $\beta$ derived in the generalized thermodynamics framework can also be considered as a driving force for exchange of the corresponding physical quantities. In the context of a large number, $M$ i.i.d. measurements, it suffices to say that when conditioned on an observation that $\overline{g}_M=y$, there is a posterior distribution $f_{\vx}(\vx)e^{-\beta(y) g(\vx)}$ among the $M$ samples \cite{vancampenhout}. If one has two sets of samples with $\overline{g}^{(1)} = y_1$ and $\overline{g}^{(2)}=y_2$, then the joint samples that pooled two together will have $\min\{y_1,y_2\} \le \overline{g}^{(1\cup 2)} \le \max\{y_1,y_2\}$.  Since $\beta(y)$ is a monotonic function of $y$, this also implies a new $\beta$ that is between the $\beta_1$ and $\beta_2$. Thus one can find the spontaneous direction of equilibration using $\vbeta$ as it is done in traditional thermodynamics. It has been shown recently that canonical distribution can also be understood as a limit theorem of an infinitesimal portion of a large system \cite{cqz}, which gives the canonical distribution and its $\beta$ an alternative interpretation.

{\bf\em Integral and differential $\vbeta$'s.}
It is worth pointing out that the conjugate force can have two distinct definitions. Consider again $K=1$ with both $\vy$ and $\vbeta$ being scalars. Then (\ref{massieu}) immediately suggests that 
\begin{eqnarray}
		-\frac{\rd\eta(y)}{\rd y} +\frac{\eta(y)}{y}
	   = -\beta + \frac{\eta(y)}{y} =  \frac{\psi}{y}.
\label{int-dif}
\end{eqnarray}
If we take the thermodynamic limit, both $\eta(y)$ and $y$ are extensive quantities, as both tending to $\infty$ if their ratio $\eta(y)/y$ exists. Then according to l'H\^{o}spital's rule the $\rd\eta(y)/\rd y$ will have the same limit as $y\to\infty$. Thus, in the thermodynamic limit the right-hand-side of (\ref{int-dif}) vanishes ($\psi$ is sub-extensive). In other words, the integral and differential definitions $\beta$ are equivalent to each other in the thermodynamic limit. In \cite{hill-book} Hill has introduced a notation $\eta(y)/y=\hat{\beta}$, called integral form of $\beta$ in contrast to the differential $\beta=\partial\eta(y)/\partial y$. Then
\begin{equation}
     -\beta+\hat{\beta} = \frac{\psi}{y} = \frac{\eta(y)-y\eta'(y)}{y}
          = -\frac{\rd\hat{\beta}}{\rd\ln y}.
\label{eq12}
\end{equation}
This is another key result in Hill's thermodynamics of small systems \cite{hill-book,bedeaux}.  The right-hand-side of (\ref{eq12}) has a very clear meaning.  To illustrate we take $y$ as the number of particles $N$ for example, then
\[
        \frac{\psi}{N} 
                =-N\frac{\rd}{\rd N}\left(\frac{\eta}{N}\right),
\]
in which $\eta/N$ is the entropy per particle.  
Accordingly $\rd(\eta/N)/\rd N$ is its change due to introducing one additional particle, and $N\rd(\eta/N)/\rd N$ is the change in the entire system, of all
$N$ particles, due to one additional particle.  In other words, when introducing one additional particle into a system of $N$ particles, the effect is ``subdivided'' into all $N$ particles.

{\bf\em Three entropies and two limits.} 
In contrast to the
classical thermodynamics with one entropy and one limit, the present theory is about three entropies and two limiting processes.  In addition to Gibbs entropy $\eta$ and free entropy $\psi$, there is a third $\ln\Omega(\vy)$ 
corresponding to the prior probability
density function for the observable $\vg(\vx)$,
\begin{eqnarray}
  \Omega(\vy)\rd\vy &=& \Pr\big\{
          \vy < \vg(\vx) \le \vy+\rd\vy\big\}
\nonumber\\
       &=& \int_{\vy<\vg(\vx)\le \vy+\rd\vy}
                          f_{\vx}(\vx)\rd\vx,
\end{eqnarray}
which completely determines the partition function and
$\psi(\vbeta)$,
\begin{equation}
     \psi(\vbeta) = \ln \int_{\mathbb{R}^K} 
         \Omega(\vy) \exp\big(-\vbeta\cdot\vy\big)\rd\vy.
\end{equation}
$\Omega(\vy)$ is the {\em density of state} in terms of
the observable $\vg(\vx)$.  $\ln\Omega$ can be identified 
as Boltzmann's microcanonical entropy if $\vg=(U,V,N)$.  In 
an essence our theory has a logic flow captured by the 
following scheme:
\begin{subequations}
\label{3ent}
\begin{eqnarray}
       \ln\Omega(\vy) &\longrightarrow& \Big\{ \psi(\vbeta)
                \leftarrow\hspace{-0.1cm}\rightarrow 
                \eta(\vy)  \Big\} 
\\
		&\longrightarrow&  
             \left\{ \frac{\psi(\vbeta)}{\vy}=0,\ 
            \frac{\eta(\vy)}{y_i}=\frac{\rd\eta}{\rd y_i}=\vbeta_i \right\}.
\end{eqnarray}
\end{subequations}
If $\ln\Omega(\vy)$ is convex, then $\ln\Omega(\vy)=\eta(\vy)$.
In this case, the microcanonical description in terms of
Boltzmann's entropy with $\vy$ as independent variables is 
equivalent to the canonical description in terms of Massieu-Guggenheim
entropy with $\vbeta$ as independent variables \cite{touchette-2}. The $y_i$ in (\ref{3ent}b) can be any one component of the extensive $\vy$. The derivative is understood as expressing all other components of $\vy$ normalized by the $y_i$.
The ``$\rightarrow$'' in (\ref{3ent}a) represents the 
repeated-measurement limit; in general there is a loss of 
information on the small 
system.  The ``$\leftrightarrow$'' in (\ref{3ent}a)
indicates the emergent duality symmetry.  This is the domain 
of Hill's nanothermodynamics \cite{hill-book}. The 
``$\rightarrow$'' in (\ref{3ent}b) then represents the large system, macroscopic thermodynamics limit. It results in a breaking of
the duality symmetry.

 This work builds on the large deviation theory to construct a unified ``probability foundation'' of Gibbs' and Hill's theories, and leads to their generalization to systems beyond thermodynamic equilibrium:  It is an emergent behavior of ``certain quantity'' becoming large: be it large measurement time, large number of particles, or large number of replica. This unified understanding generalizes the thermodynamic limit into a measurement limit and is clearly beyond the work of Hill, or Gibbs theory itself. The thermodynamic structure presented in the present work,
while assumes a probability distribution {\it a priori}, does not 
require the concept of {\em equilibrium} in connection to 
detailed balance in stochastic dynamics, nor {\em ergodicity}.
Therefore, it is applicable to measurements on biomarkers from
isogenic single living cells \cite{qian-cheng-qb}.   Of course, if 
a large system consists of many statistically identical but 
independent smaller parts, then the entire argument based 
on i.i.d. measurements can be applied to a single measurement 
of extensive variables of the large system as a whole.  This
mathematics was precisely in Boltzmann's theory of 1877, with 
the probabilistic concept of ``conditioning on the value $y$'' being
replaced by Newtonian ``conservation with the value $y$''. 
 
	The present result augments the current understanding of 
the nature of thermodynamic behavior, which so far has been 
focused on large systems limit and phase transition through 
symmetry breaking as a key route for emergent phenomena \cite{pwanderson,qcy-20}.  We now see there is actually a 
{\em large measurements limit} that generates a different kind of emergent order, a duality symmetry, for any small stochastic systems with stationarity or replica.  This symmetry is lost, however, in the large systems limit.

%{\bf\em free entropy of a fluctuating entropy?} 
% The new theory might also help in further investigating some 
% of the standing issues in the field of statistical thermodynamics.
% Take a discrete system with probability mass function 
% $\Pr\{X=k\}=p(k)$ for example, it has been shown recently 
% that $\ln p(X)$, as a function of the random variable 
% $X$, is the {\em fluctuating entropy} in stochastic 
% thermodynamics \cite{qct}.  Then if one constructs the 
% free entropy $\psi(\beta)$ corresponding to the random 
% variable $\ln p(X)$:
% \begin{equation}
%      \psi(\beta) = \ln \sum_{k} p(k) e^{-\beta
%       \ln p(k)}  = \ln
%         \sum_{k} p^{1-\beta}(k),
% \end{equation}
% which looks like the Tsallis entropy.  In fact, 
% if $\beta=1-q\ll 1$, one has the corresponding 
% negative free energy
% \begin{equation}
%   \frac{\psi(\beta) }{\beta}=  \frac{1}{1-q}
%       \left(\sum_{k} p^{q}(k)-1\right).
% \label{tsallis}
% \end{equation}

{\bf\em Acknowledgements.}
Z. L. acknowledges the financial support from the startup fund at UNC-Chapel Hill. H. Q. is grateful to Prof. Lu Yu (ITP, CAS) for advice and the Olga Jung Wan Endowed Professorship. We thank D. Bedeaux, R. Chamberlin, C. Jarzynski, S. Kjelstrup, the members of Summer 2020 Online Club Nanothermodynamica, 
especially Quanhui Liu and Hong Zhao, for extensive discussions, and E. Smith, A. Szabo, P. Warren for comments. We appreciate the reviewers for constructive suggestions on the presentation of this work.

\bibliographystyle{apsrev4-1}
\bibliography{apssamp}% Produces the bibliography via BibTeX.

\end{document}